\documentclass[prl,aps,floats,floatfix,superscriptaddress,twocolumn,showpacs]{revtex4}

\usepackage[dvips]{graphicx}
\usepackage{amsmath} 
\usepackage{longtable}
\usepackage{dcolumn,epsfig}

\def\laq{\raise 0.4ex\hbox{$<$}\kern -0.8em\lower 0.62ex\hbox{$\sim$}}
\def\gaq{\raise 0.4ex\hbox{$>$}\kern -0.7em\lower 0.62ex\hbox{$\sim$}}

\newcommand{\beq}{\begin{equation}}
\newcommand{\eeq}{\end{equation}}
\newcommand{\bea}{\begin{eqnarray}} 
\newcommand{\eea}{\end{eqnarray}}
\newcommand{\ba}{\begin{array}}
\newcommand{\ea}{\end{array}}

\newcommand{\comment}[1]{}

\newlength{\sizeonefig}
\newlength{\sizetwofig}
\newlength{\sizeonefigb}
\newlength{\sizetwofigb}
\begin{document}
\title{Displacement-Noise-Free
Gravitational-Wave Detection
}
\author{Seiji Kawamura}
\affiliation{TAMA Project, National Astronomical Observatory of Japan, Mitaka, Tokyo,
Japan}
\author{Yanbei Chen}
\affiliation{Theoretical Astrophysics, California Institute of Technology, Pasadena, CA 91125}


\begin{abstract}
We present a  new idea that allows us to detect gravitational
waves without being disturbed by any kind of displacement noise, based on the fact that gravitational waves 
and test-mass motions affect the propagations of light differently. We 
demonstrate this idea by analyzing a simple toy model consisting three equally-separated objects 
on a line. By taking a certain combination of light 
travel times between these objects, we construct an observable free from 
the displacement of each object, which has a reasonable sensitivity to 
gravitational waves.  
\end{abstract}
\maketitle


Gravitational waves (GWs) have been predicted by Einstein's general relativity~\cite{GW}, but have not yet been detected directly. Many attempts have been (or are being) made to detect them, including resonant-type detectors~\cite{bars}, Doppler tracking~\cite{Doppler}, pulsar timing~\cite{pulsar}, and laser interferometric detectors, both ground-based~\cite{Laser} and space-based~\cite{Laserspace}.  Since GW
signals are very weak, any undesirable motion of test objects used in a
detector could prevent the detection. Here we present a new idea that  allows us to detect GWs without being contaminated by motions of test objects.

The effect of GWs with wavelength $\lambda$,  in the proper reference frame of an observer~\cite{GW}, for test particles within  proper distance $L \ll \lambda $, is simply equivalent to a time-dependent tidal-force field, whose amplitude is proportional to the distance from the observer, as well as the mass of test particles (so the induced acceleration is independent of mass). The accuracy of this approximation is of the order  $\mathcal{O}(2\pi L/\lambda)$.  The effect of GWs on light propagation is of an even higher order, $\mathcal{O}\left[(2\pi L/\lambda)^2\right]$. 
Within this approximation, any kind of (non-geodesic) test-mass motion induced by external forces will be indistinguishable from GW signals, and become a {\it displacement noise} of the detector. 

If the size of an instrument is comparable to the wavelength, however, the above approximation is no longer valid, and it becomes easiest to analyze the instrument in the so-called TT coordinate system, in which freely falling test masses have fixed spatial coordinates, while the effect of GWs can be thought of as an apparent change in the (coordinate) speed of the light. As a consequence, even if the test masses are not ideal and move around in this coordinate system, signals in GW detectors respond to GWs differently from test-mass motions.  Indeed, as we shall demonstrate,  this difference can be used to separate GWs from  test-mass motions, and hence be used to construct {\it displacement-noise-free} GW detectors.

\begin{figure}[b]
\vspace{-0.2cm}
\centerline{\includegraphics[width=0.3\textwidth]{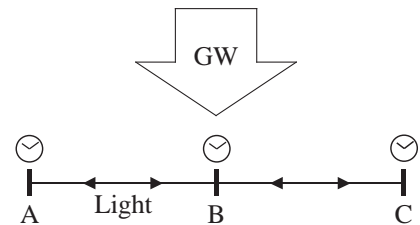}}
\caption{Schematic view of the configuration of three objects (A, B, and C) and GW propagating perpendicularly to the line consisting of the three objects. \label{fig1}}
\end{figure}

\begin{figure}[t]
\begin{center}
\centerline{\includegraphics[width=0.475\textwidth]{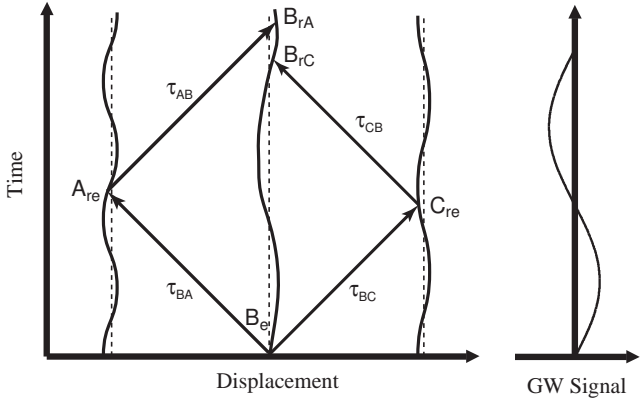}}
\end{center}
\vspace{-1cm}
\caption{Simplified illustration of how to measure GWs without being contaminated by the displacement noise. \label{fig2}}
\end{figure}

Let us consider three freely falling objects, $A$, $B$, and $C$, equally separated on a line, 
and a GW propagating perpendicularly to this line  (Fig.~\ref{fig1}) with  period  equal to the round-trip time of light between $A$ and $B$ (or  $B$ and $C$) (these assumptions will be relaxed later). [This configuration is remarkably similar to the SyZyGy scheme proposed by Tinto, Estabrook and Armstrong~\cite{syzygy} for space-based GW detection, but the aim here is very different. Different quantities will be measured in our toy model.]  Each object also carries a clock, and all three clocks are synchronized perfectly to each other in the absence of GWs. Now each object is allowed to be moving randomly by a small amount for various reasons.  As shown in Fig.~\ref{fig2}, we assume that light pulses are emitted from $B$ toward both $A$ and $C$ simultaneously.  The time of emission is measured by clock $B$. Let $B_e$ be the displacement of $B$ at the time of emission. The two pulses emitted from $B$ reach $A$ and $C$, at times of arrival measured by clocks $A$ and $C$, respectively. The pulses are promptly reflected from $A$ and $C$ toward $B$ without any delay. Let $A_{re}$ and $C_{re}$ be displacements of A and C at the times of reflection, respectively. The two pulses then return to $B$, with times of arrival measured by clock $B$. Let $B_{rA}$ and $B_{rC}$
be the displacements of $B$ at these two instants. Since the difference between the two times of arrival is of order $\mathcal{O}(hL/c)$ (where $h$ is the GW amplitude), the difference between  $B_{rA}$/c and  $B_{rC}/c$ is of order $(v/c) (hL/c)$, and can be ignored when $v\ll c$, where $v$ is the speed of mass $B$.

Now we can obtain the light travel time, as measured by the three ideal clocks $\tau_{\mbox{\tiny BA}}$ (from $B$ to  $A$), $\tau_{\mbox{\tiny AB}}$ (from $A$ back to $B$), $\tau_{\mbox{\tiny BC}}$ (from $B$ to $C$), $\tau_{\mbox{\tiny CB}}$ (from $C$ back to $B$) :
\bea
\label{tauba}
\tau_{\mbox{\tiny BA}}&=&+S_{g}+(B_{e}-A_{re})/c\,, \\
\tau_{\mbox{\tiny AB}}&=&-S_{g}+(B_{rA}-A_{re})/c\,, \\
\tau_{\mbox{\tiny BC}}&=&+S_{g}+(C_{re}-B_{e})/c\,, \\
\tau_{\mbox{\tiny CB}}&=&-S_{g}+(C_{re}-B_{rC})/c\,. 
\label{taucb}
\eea
Here $c$ is the speed of light, and $S_{g}$ is the GW signal.  Now let us consider the following quantity:
\beq
\label{tau0}
\tau_{0} = \tau_{\mbox{\tiny BA}} - \tau_{\mbox{\tiny AB}}+\tau_{\mbox{\tiny BC}}-\tau_{\mbox{\tiny CB}}\,.
\eeq
Inserting Eqs.~\eqref{tauba}--\eqref{taucb} into Eq.~\eqref{tau0}, with the approximation that $B_{rA} \approx B_{rC}$, we obtain
$\tau_{0}=4 S_g$.

The use of linear combinations of timing signals makes our scheme look
similar to the Time-Delay Interferometry (TDI)  proposed for
LISA~\cite{TDI}; the aims, however, are different  (if complementary): TDI
operates analogous to a conventional interferometric detector, which
cancels the laser noise (or clock noise in our context), while retaining the
displacement noise. For example, by contrast to Eq.~\eqref{tau0}, the
combinations $\tau_{\mbox{\tiny BA}}+\tau_{\mbox{\tiny AB}}$ and
$\tau_{\mbox{\tiny BC}}+\tau_{\mbox{\tiny CB}}$ are used by TDI in order to
avoid laser noises from $A$ and $C$;  the combinations $\tau_{\mbox{\tiny
BA}}-\tau_{\mbox{\tiny BC}}$ and $\tau_{\mbox{\tiny AB}}-\tau_{\mbox{\tiny
CB}}$ are used to cancel laser noise from $B$. Moreover, TDI does not provide the insight that GWs and test-mass motions affect the
propagations of light differently. In this sense, the idea of the
displacement-noise-free GW detection is very different from
that of TDI.

Our scheme is possible also from the following consideration. In Eqs.~\eqref{tauba}--\eqref{taucb} we have four independent measured values, four unknown displacement-noise terms ($B_e$, $A_{re}$, $C_{re}$ and $B_{rA} \approx B_{rC}$), and the GW signal. However the four unknown displacement-noise terms can be reduced into three (e.g., $A_{re}-B_e$, $C_{re}-B_e$ and $B_{rA}-B_e \approx B_{rC}-B_e$), since only relative differences between them enter the equations. In this way, we have four equations, and only three unknown displacement-noise terms, so it must be possible to cancel all of them and extract some GW signal. 
\begin{figure}[b]
\vspace{-0.1cm}
\begin{center}
\centerline{\includegraphics[width=0.275\textwidth]{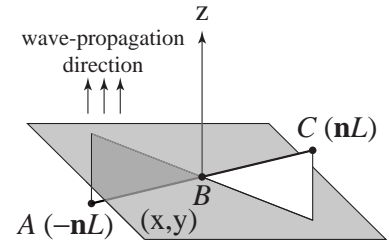}}
\end{center}
\vspace{-0.75cm}
\caption{Ideal positions of test masses $A$, $B$, and $C$ in a 3-surface of constant $t$.
 \label{fig3}}
\end{figure}

A rigorous derivation of our device's response to GWs and hence its insusceptibility to displacement noise is rather straightforward;
the method of derivation is well-known, as given, for example, by Estabrook and Wahlquist~\cite{EW}, and used by Tinto, Estabrook and Armstrong~\cite{syzygy}.  However, in order to demonstrate our idea rigorously, we present this elementary derivation here once more for completeness.
We adopt the TT coordinate system, in which  a weak GW is propagating in the $+z$ direction  (we shall set $c=1$ here and through the rest of the paper): 
\bea
ds^2 =& -&dt^2+dz^2+\big[1+h_+(t-z)\big]dx^2 \nonumber \\
                 &+&2h_{\times}(t-z)dx dy 
      + \big[1-h_+(t-z)\big]dy^2\,.
\eea
In this coordinate system, free test masses that start with vanishing (coordinate) speed will stay static, i.e., the world line described by $x_{\rm static}^\mu =(t,x,y,z)$, with $t\in(-\infty,+\infty)$ and constant $(x,y,z)$ is a timelike geodesic.
Suppose our ideal detector consists of three such freely falling objects, $A$, $B$ and $C$ (our test masses). In addition, in the 3-surface of constant $t$, $AB$ and $BC$ both lie in some generic direction
\beq
\mathbf{n} = (n_x,n_y,n_z)\,,\quad n_x^2+n_y^2+n_z^2=1\,,
\eeq
and have coordinate length  $L$ (see Fig.~\ref{fig3}). In reality, the test masses will have the following world lines
\bea
A:  \big[t,\mathbf{x}_A(t)-\mathbf{n}L\big]\;\,
B: \big[t,\mathbf{x}_B(t)\big]\;\,
C: \big[t,\mathbf{x}_C(t)+\mathbf{n}L\big],\nonumber
\eea
with $\mathbf{x}_{A,B,C}(t)$ displacement noises, which are going to be treated as small quantities, in the following sense: (i) $|\mathbf{x}_{A,B,C}(t)| \ll L$, and (ii) $|\dot{\mathbf{x}}_{A,B,C}(t)| \ll  1$, where $\dot{\mathbf{x}} \equiv d\mathbf{x}/dt$.
The proper times of observers traveling together with our test masses, as indicated by ideal clocks sitting on them, deviate from the coordinate time $t$ only at second order in $\dot{\mathbf{x}}_{A,B,C}(t)$, and will be ignored all through our analysis.

\begin{figure}[b]
\vspace{-0.1cm}
\begin{center}
\centerline{\includegraphics[height=0.26\textwidth]{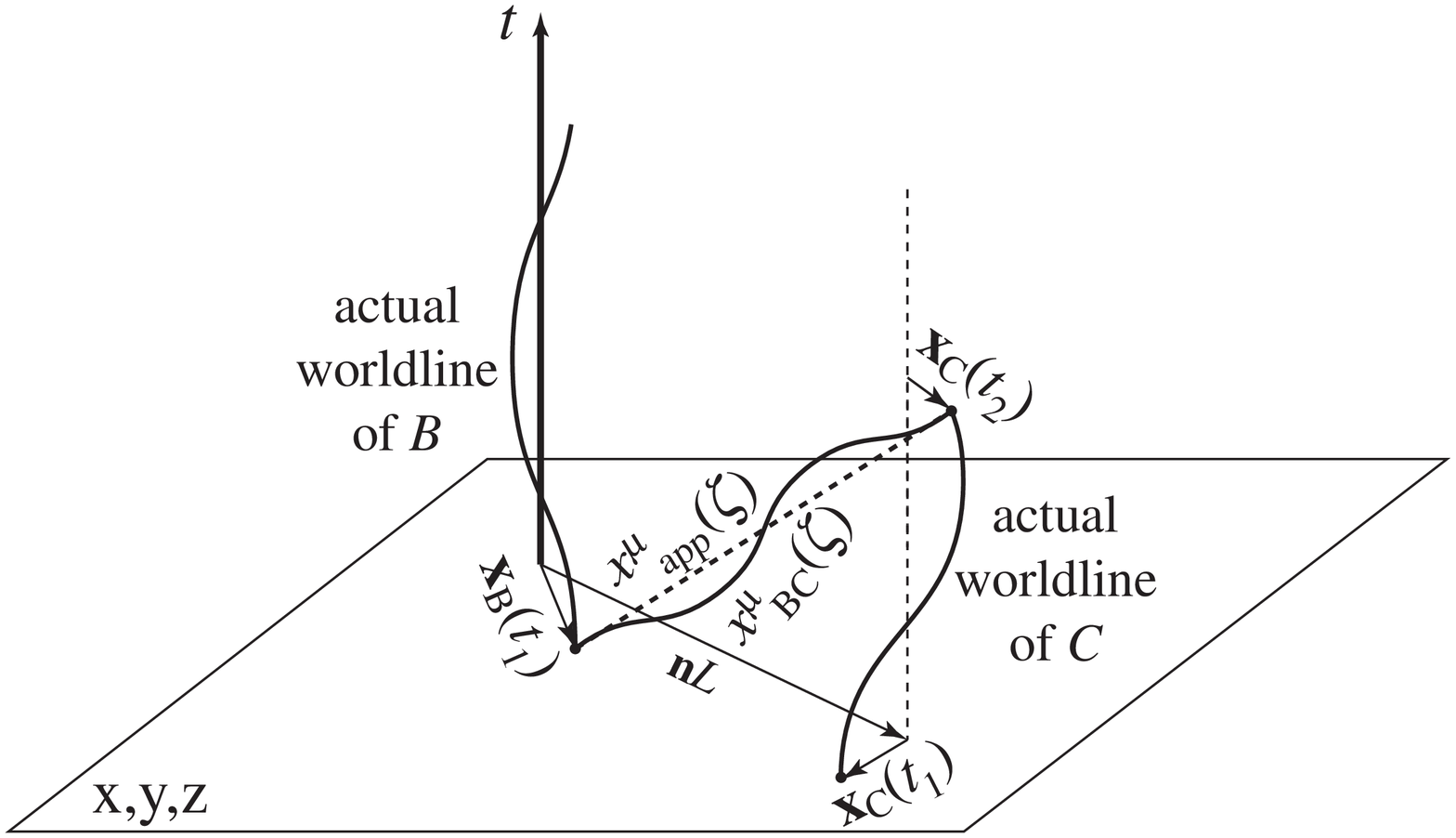}}
\end{center}
\vspace{-0.75cm}
\caption{Schematic plot showing the propagation of a pulse from $B$ to $C$. In this figure, we use a 2-D plane to illustrate the constant-$t$ surface with $t=t_1$, the pulse emission time; spatial positions of $B$ and $C$ at this instant (as measured by coordinate time), $\mathbf{x}_B(t_1)$ and $\mathbf{n}L+\mathbf{x}_C(t_1)$, are illustrated in this plane as 2-D vectors. Ideal world lines of $B$ and $C$ are the $t$ axis and the dashed line parallel to it, respectively, while actual world lines, which differ from them by a small amount, are illustrated by two curves around them.  The null geodesic from $B$ to $C$, $x^\mu_{BC}(\zeta)$, along which the light pulse travels, is shown as a solid curve connecting the emitting and receiving points, while the approximated path  $x^\mu_{\rm app}(\zeta)$ is shown as a dashed line connecting them.  
 \label{fig4}}
\end{figure}

Now suppose that a light pulse starts from $B$ at $t_1$, i.e., $\big[t_1,\mathbf{x}_B(t_1)\big]$, and reaches $C$ at $t_2$, i.e., 
$\big[t_2,\mathbf{x}_C(t_2)+\mathbf{n}L\big]$ (see Fig.~\ref{fig4}), following a null geodesic $x_{BC}^\mu(\zeta)$, with an affine parameter $\zeta$ running from 0 to 1 (solid curve from $B$ to $C$ in Fig.~\ref{fig4}).  
It is straightforward to argue that, the world line 
\bea
\label{appwl}
&&x_{\rm app}^\mu(\zeta)\nonumber\\
&=& \Big[t_1+(t_2-t_1)\zeta, 
\nonumber \\
&&\qquad\qquad
\mathbf{x}_B(t_1)+\left[\mathbf{n}L+\mathbf{x}_C(t_2)-\mathbf{x}_B(t_1)\right]\zeta\Big]\,,
\eea
which has the same ends as $x_{BC}^\mu(\zeta)$, is $\mathcal{O}(h)$ away from it for any $0<\zeta<1$ (dashed line from $B$ to $C$ in Fig.~\ref{fig4}). Since the action
\beq
I[x^\mu(\zeta)]=\int_0^1  g_{\mu\nu}
\frac{dx^\mu}{d\zeta}\frac{dx^\nu}{d\zeta}d\zeta
\eeq
has an extremum of $0$ at $x_{\rm BC}^\mu(\zeta)$, we have
\beq
\label{Iapp}
I[x_{\rm app}^\mu(\zeta)]=\mathcal{O}(h^2)\,.
\eeq
On the other hand, for the LHS of Eq.~\eqref{Iapp}, we have
\bea
I[x_{\rm app}^\mu(\zeta)] &=& -2  (t_2-t_1-L) L \nonumber \\
&+&2 L \mathbf{n}\cdot \left[\mathbf{x}_C(t_1+L)-\mathbf{x}_B(t_1)\right]\nonumber \\
&+&  2 L \sum_{p=+,\times} F_p\int_0^L h_p\left[t_1+(1-n_z)t'\right]dt' \nonumber \\
&+& \mathcal{O}\left[h^2,\,h x_{B,C}^i,\,x_{B,C}^i x_{B,C}^j\right]\,,
\eea
where
\beq
F_+=({n_x^2-n_y^2})/{2}\,,\qquad F_\times=n_xn_y\,,
\eeq
and 
\beq
\label{FpFx}
\sqrt{ F_+^2 + F_\times^2} =({1-n_z^2})/{2}\,.
\eeq
It is then straightforward to obtain, up to leading order in $h$ and $\mathbf{x}_{B,C}$:
\bea
\label{tauBC}
\tau_{\mbox{\tiny BC}} = L &+& \mathbf{n}\cdot \left[\mathbf{x}_C(t_1+L)-\mathbf{x}_B(t_1)\right]  \nonumber \\
&+& \sum_{p=+,\times} F_p \int_0^L h_p\left[t_1+(1-n_z)t'\right]dt'\,. \quad
\eea
(Here we recall the agreement between coordinate time and proper time up to linear order in test-mass motion.) Similarly, for the pulse relayed back from $C$ to $B$, we have
\bea
\label{tauCB}
&&\tau_{\mbox{\tiny CB}}\nonumber \\
&=& L + \mathbf{n}\cdot \left[\mathbf{x}_C(t_1+ L)-\mathbf{x}_B(t_1+2 L)\right]  \nonumber \\
&+& \sum_{p=+,\times} F_p  \int_0^L h_p\left[t_1+(1-n_z)L+(1 + n_z)t'\right]dt' \,,
\qquad
\eea
in which $t_2$ has been  replaced by $t_1+L$, which gives the correct answer up to leading order in $h$ and $\mathbf{x}_{B,C}$. Subtracting Eq.~\eqref{tauBC} from Eq.~\eqref{tauCB}, we obtain
\bea
\label{CBmBC}
&&\tau_{\mbox{\tiny CB}}-\tau_{\mbox{\tiny BC}}\nonumber \\
&=&\mathbf{n}\cdot \left[\mathbf{x}_B(t_1)-\mathbf{x}_B(t_1+2L)\right] \nonumber \\
&+& \sum_{p=+,\times} F_p
\int_0^L dt' \big\{h_p\left[t_1+(1-n_z)L+(1+n_z)t'\right] \nonumber \\
&& \qquad\qquad\qquad\;\; -h_p\left[t_1+(1-n_z)t'\right]\big\}\,.
\qquad\qquad\qquad
\eea 
For the pulse sent to $A$ and relayed back, we take Eq.~\eqref{CBmBC} and apply $C\rightarrow A$, $\mathbf{n}\rightarrow -\mathbf{n}$, obtaining
\bea
\label{ABmBA}
&&\tau_{\mbox{\tiny AB}}-\tau_{\mbox{\tiny BA}}\nonumber \\
&=&-\mathbf{n}\cdot \left[\mathbf{x}_B(t_1)-\mathbf{x}_B(t_1+2L)\right] \nonumber \\
&+& \sum_{p=+,\times} F_p
\int_0^L dt' \big\{h_p\left[t_1+(1+n_z)L+(1-n_z)t'\right] \nonumber \\
&& \qquad\qquad\qquad\;\; -h_p\left[t_1+(1+n_z)t'\right]\big\}\,.
\qquad\qquad\qquad
\eea 
Assembling Eqs.~\eqref{CBmBC} and \eqref{ABmBA}, we obtain
\beq
\label{tau0_math}
\tau_0= (\tau_{\mbox{\tiny BC}}-\tau_{\mbox{\tiny CB}})+(\tau_{\mbox{\tiny BA}}-\tau_{\mbox{\tiny AB}})
= \sum_{p=+,\times} F_p H_p\,,
\eeq
where
\bea
H_{p}(t_1)
= \int_0^L  dt' \Big\{
\!\!\!\!&-&\!\!\!h_p\left[t_1+(1+n_z)L+(1-n_z)t'\right] \nonumber \\
\!\!\!\!&-& \!\!\!h_p\left[t_1+(1-n_z)L+(1+n_z)t'\right]\nonumber \\
\!\!\!\!&+& \!\!\! h_p\left[t_1+(1-n_z)t'\right] \nonumber \\
\!\!\!\!&+&\!\!\! h_p\left[t_1+(1+n_z)t'\right]
\Big\}\,,\; p=+,\times,\quad\quad
\eea
from which the displacement noises have been removed. 
Here we note that, the argument $t_1$ in $H_p$ represents the fact that this time combination is obtained via pulses sent from $B$ at time $t_1$. 
The pulses are received back at B at a later time $\approx t_1+2L$. In the frequency domain, we have
\bea
\tilde{H}_p(\Omega)
\!&=&\!(L\tilde{h}_p)(2 i\Omega L) e^{-i\Omega L}{\rm sinc}(\Omega L(1-n_z)/2)\nonumber \\
&&\quad\quad \qquad \qquad \;\;\;\;\times{\rm sinc}(\Omega L(1+n_z)/2)\,,\quad\;
\eea
where $\mathrm{sinc}(x)\equiv x^{-1} \sin x$.  

\begin{figure}[t]
\begin{center}
\centerline{\includegraphics[height=0.32\textwidth]{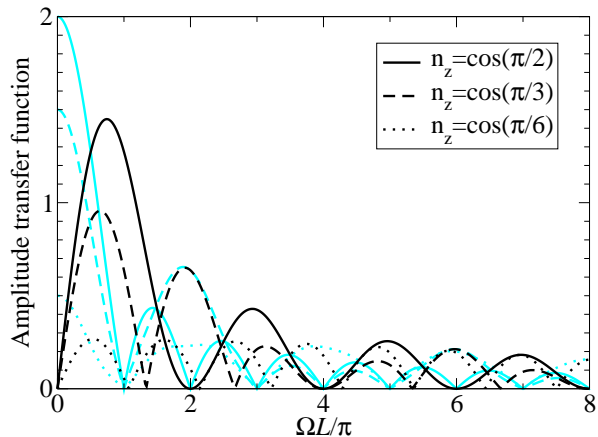}}
\end{center}
\vspace{-0.7cm}
\caption{Amplitude transfer functions [Eq.~\eqref{transf}] from GWs to $\tau_0$, when the detector line $ABC$ is  $90^\circ$ ($n_z=\cos\pi/2$, dark solid curve), $60^\circ$ ($n_z=\cos\pi/3$, dark dashed curve) and $30^\circ$ ($n_z=\cos\pi/6$, dark dotted curve) from the wave propagation direction. Also shown for comparison (light curves) are transfer functions from GWs to $\tau_{\mbox{\tiny sym}}$, for the same set of detector orientations.   
 \label{fig5}}
\end{figure}

In Fig.~\ref{fig5}, we plot the transfer function from $h$ to $\tau_0$,  
\beq
\label{transf}
\mathcal{T}(\Omega) \equiv \left| [c\tilde \tau_0(\Omega)]/[L\tilde{h}(\Omega)] \right|\,,
\eeq 
 for cases when the detector line $ABC$ is $90^\circ$ ($n_z=\cos\pi/2$, dark solid curve), $60^\circ$ ($n_z=\cos\pi/3$, dark dashed curve) and $30^\circ$ ($n_z=\cos\pi/6$, dark dotted curve) from the wave propagation direction. For simplicity, we assume the wave to be purely $+$-polarized, and have our detector lie inside the $x$-$z$ (or $x$-$y$) plane, so that we have $F_+=(1-n_z^2)/2$ [which is optimal for detecting this polarization, see also Eq.~\eqref{FpFx}.] Note that, our detector is most sensitive to GWs that (i) propagate perpendicular to the line $ABC$, or (ii) have frequency around a half-odd-number times $c/L$. By contrast, GWs propagating along the $ABC$ direction, or with period equal to an integer times $c/L$ ({\it including dc}),  cannot be detected. For comparison, we also plot (in light curves) the transfer function of a more conventional combination,
 \beq 
 \label{taumich}
 \tau_{\mbox{\tiny sym}} \equiv 
 (\tau_{\mbox{\tiny CB}}+\tau_{\mbox{\tiny BC}})+(\tau_{\mbox{\tiny AB}}+\tau_{\mbox{\tiny BA}}),
  \eeq
which has a {\it maximal} response to GWs at DC.  As we can see from the figure, except for frequencies near DC, $\tau_0$ [Eqs.~\eqref{tau0} and \eqref{tau0_math}] has comparable responses to GWs to  $\tau_{\mbox{\tiny sym}}$ [Eq.~\eqref{taumich}]. 

In this way, we have demonstrated, using a simple toy model, that the difference in the ways GWs and test-mass motions influence the propagations of light allows displacement-noise-free detection of GWs at all frequencies except DC; peak sensitivities of our model detector lies roughly at odd multiples of $c/(2L)$~\footnote{For ground-based detectors, $c/(2L)$ is usually much higher than frequencies of currently conceivable GWs (below 10\,kHz). At these lower frequencies, our scheme still evades the displacement noise, but our signal strength can be much lower than more conventional schemes, e.g.,  the one given by Eq.~\eqref{taumich}. }. In particular, our model detector is not susceptible to the radiation-pressure noise, so the quantum noise can be lowered indefinitely, even below the standard quantum limit, by increasing the light power\footnote{As pointed out by V.~Braginsky, the device's back actions to the GW being measured will eventually limit the accuracy of our scheme.}.  In our toy model, we have used a linear geometry, but this is not essential: for example, a similar, displacement-noise-free scheme can also be devised for an L-shaped configuration.  
Finally, we speculate that this idea may eventually lead us to a significant improvement in the sensitivity of gravitational wave detectors.

We would like to thank V.~Braginsky, Y.~Levin, T.~Nakamura, M.~Sasaki, N.~Seto, K.~Somiya, H.~Tagoshi, M.~Tinto, K.~Thorne, M.~Vallisneri and S.~Whitcomb for useful discussions on this subject. The research of Y.C is supported by NSF grant PHY-0099568 and by the David and Barbara Groce Fund at the San Diego Foundation.



\end{document}